%% file: menu07-zoubs0.tex
\documentclass[12pt,a4paper]{conference}

\usepackage{fancyhdr}
\usepackage{graphicx,amsmath,amssymb,cite}
\usepackage{multind}
\makeindex{author} \makeindex{subject}

\input MENU07macros.tex

\begin{document}

\Chapter{Baryon Resonances Observed at BES}
           {Baryon Resonances Observed at BES}{B. S. Zou}
\vspace{-6 cm}\includegraphics[width=6 cm]{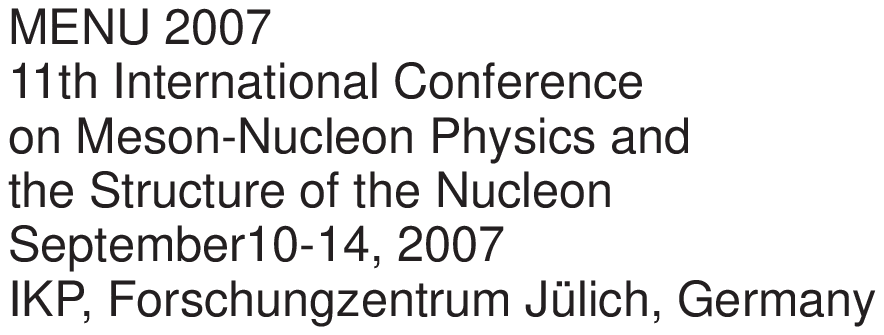}
\vspace{4 cm}

\addcontentsline{toc}{chapter}{{\it N. Author}} \label{authorStart}

\begin{raggedright}

{\it B. S. Zou}\index{author}{Zou, B. S.}\\
Institute of High Energy Physics, Chinese Academy of Sciences, \\
P.O.Box 918(4), Beijing 100049, China
\bigskip\bigskip

\end{raggedright}

\begin{center}
\textbf{Abstract}
\end{center}
The $\psi$ decays provide a novel way to explore baryon spectroscopy
and baryon structure. The baryon resonances observed from $\psi$
decays at BES are reviewed.  The implications and prospects at
upgraded BESIII/BEPCII are discussed.

\section{Introduction}

Although the quark model achieved significant successes in the
interpretation of a lot of static properties of nucleons and the
excited resonances, our present knowledge on baryon spectroscopy is
still in its infancy \cite{PDG}. Many fundamental issues in baryon
spectroscopy are still not well understood \cite{Capstick1}.

\begin{figure}[ht]
\vspace{0cm}
\hspace{0cm}\includegraphics[scale=0.4]{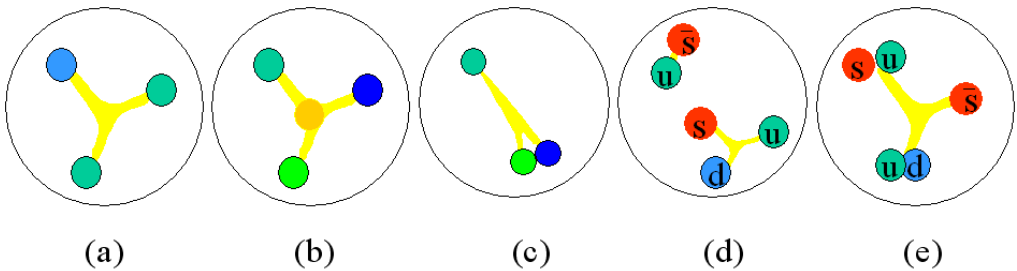}
\caption{Various pictures for internal quark-gluon structure of
baryons: (a) $3q$, (b) $3qg$ hybrid, (c) diquark, d) meson-baryon
state, (e) pentaquark with diquark clusters. } \label{fig1}
\end{figure}

On theoretical side, an unsolved fundamental problem is: what are
proper effective degrees of freedom for describing the internal
structure of baryons? Several pictures based on various effective
degrees of freedom are shown in Fig.\ref{fig1}.

The classical simple 3q constituent quark model as shown by
Fig.\ref{fig1}(a) has been very successful in explaining the static
properties, such as mass and magnetic moment, of the spatial ground
states of the flavor SU(3) octet and decuplet baryons. Its predicted
$\Omega$ baryon with mass around 1670 MeV was discovered by later
experiments. However its predictions for the spatial excited baryons
are not so successful. In the simple 3q constituent quark model, the
lowest spatial excited baryon is expected to be a ($uud$) $N^*$
state with one quark in orbital angular momentum $L=1$ state, and
hence should have negative parity. Experimentally \cite{PDG}, the
lowest negative parity $N^*$ resonance is found to be $N^*(1535)$,
which is heavier than two other spatial excited baryons :
$\Lambda^*(1405)$ and $N^*(1440)$. In the classical 3q constituent
quark model, the $\Lambda^*(1405)$ with spin-parity $1/2^-$ is
supposed to be a ($uds$) baryon with one quark in orbital angular
momentum $L=1$ state and about 130 MeV heavier than its $N^*$
partner $N^*(1535)$; the $N^*(1440)$ with spin-parity $1/2^+$ is
supposed to be a ($uud$) state with one quark in radial $n=1$
excited state and should be heavier than the $L=1$ excited ($uud$)
state $N^*(1535)$, noting the fact that for a simple harmonic
oscillator potential the state energy is $(2n+L+3/2)\hbar\omega$. So
for these three lowest spatial excited baryons, the classical quark
model picture is already failed.

The second outstanding problem in the classical 3q quark model is
that in many of its forms it predicts a substantial number of
`missing $N^*$ states' around 2 GeV/$c^2$, which have not so far
been observed \cite{Capstick1}. Since the more number of effective
degrees of freedom the more predicted number of excited states, the
`missing $N^*$ states' problem is argued in favor of the diquark
picture as shown in Fig.\ref{fig1}(c) which has less degree of
freedom and predicts less $N^*$ states \cite{diquark}. For example,
in diquark models, the two quarks forming the diquark are
constrained to be in the relative S-wave, and hence cannot combine
the third quark to form (20,$1^+_2$)-multiplet baryons.
Experimentally, not a single (20,$1^+_2$)-multiplet baryon has been
identified yet \cite{PDG}.  However, non-observation of these
`missing $N^*$ states' does not necessarily mean that they do not
exist. In the limit that the $\gamma$ or $\pi$ couples to one quark
in the nucleon in the $\gamma N$ or $\pi N$ reactions, the
(20,$1^+_2$)-multiplet baryon cannot be produced \cite{Zhaoq}.
Considering higher order effects, they may have weak coupling to
$\pi N$ and $\gamma N$, but maybe too weak to be produced by
presently available $\pi N$ and $\gamma N$ experiments
\cite{Capstick1,Zhaoq}. Other production processes should be
explored. Moreover the diquark models are only successful for very
limited aspects.

The third outstanding problem for the classical 3q quark model is
that from deep inelastic scattering and Drell-Yan experiments the
number of $\bar d$ is found to be more than the number of $\bar u$
by 0.12 in the proton \cite{Garvey}. This is argued in favor of a
mixture of the meson-baryon states as shown by Fig.\ref{fig1}(d).
With this picture, the excess of $\bar d$ over $\bar u$ in the
proton is explained by a mixture of $n\pi^+$ with the $\pi^+$
composed of $u\bar d$ \cite{Thomas}; the $N^*(1535)$ and
$\Lambda^*(1405)$ are ascribed as quasi-bound states of $K\Sigma$
and $\bar KN$, respectively \cite{Weise}. The extreme of this
picture is that only the ground state baryon-octet $1/2^+$ and
baryon-decuplet $3/2^+$ are dominated by qqq while all excited
baryons are generated by meson-baryon coupled channel dynamics
\cite{Lutz,Oset1}. However the mixture of the pentaquark components
with diquark clusters as shown by Fig.\ref{fig1}(e) can also explain
these properties \cite{zou1,zr,liubc,zhusl}.

Another possible configuration for baryons is $gqqq$ hybrid as shown
by Fig.\ref{fig1}(b) with various phenomenological models reviewed
by Ref.\cite{Barnes1}.

In reality for a baryon state around 2 GeV, it could be a mixture of
all five configurations shown in Fig.\ref{fig1}.

On experimental side, our present knowledge of baryon spectroscopy
came almost entirely from partial-wave analyses of $\pi N$ total,
elastic, and charge-exchange scattering data of more than twenty
years ago \cite{PDG}. Only recently, the new generation of
experiments on $N^*$ physics with electromagnetic probes at CEBAF at
JLAB, ELSA at Bonn, GRAAL at Grenoble and SPRING8 at JASRI have been
producing some nice results. However, a problem for these
experiments is that above 1.8 GeV there are too many broad
resonances with various possible quantum numbers overlapping with
each other and it is rather difficult to disentangle them. Moreover
resonances with weak couplings to $\pi N$ and $\gamma N$ will not
show up in these experiments.

Joining the new effort on studying the excited nucleons, $N^*$
baryons, BES started a baryon resonance program \cite{Zou1} at
Beijing Electron-Positron Collider (BEPC). The $J/\psi$ and $\psi'$
experiments at BES provide an excellent place for studying excited
nucleons and hyperons -- $N^*$, $\Lambda^*$, $\Sigma^*$ and $\Xi^*$
resonances \cite{Zou2}.

Comparing with other facilities, our baryon program has advantages
in at least three obvious aspects:

(1) We have pure isospin 1/2 $\pi N$ and $\pi\pi N$ systems from
$J/\psi\to\bar NN\pi$ and $\bar NN\pi\pi$ processes due to isospin
conservation, while $\pi N$ and $\pi\pi N$ systems from $\pi N$ and
$\gamma N$ experiments are mixture of isospin 1/2 and 3/2, and
suffer difficulty on the isospin decomposition;

(2) $\psi$ mesons decay to baryon-antibaryon pairs through three or
more gluons. It is a favorable place for producing hybrid (qqqg)
baryons, and for looking for some ``missing" $N^*$ resonances, such
as members of possible (20,$1^+_2$)-multiplet baryons, which have
weak coupling to both $\pi N$ and $\gamma N$, but stronger coupling
to $g^3N$;

(3) Not only $N^*$, $\Lambda^*$, $\Sigma^*$ baryons, but also
$\Xi^*$ baryons with two strange quarks can be studied. Many
QCD-inspired models \cite{Capstick1} are expected to be more
reliable for baryons with two strange quarks due to their heavier
quark mass. More than thirty $\Xi^*$ resonances are predicted where
only two such states are well established by experiments. The theory
is totally not challenged due to lack of data.

In this paper, we review baryon resonances observed by BESI and
BESII, and discuss the prospects for baryon spectroscopy at BESIII.

\section{Baryon Spectroscopy at BESI and BESII}

BESI started data-taking in 1989 and collected 7.8 million $J/\psi$
events and 3.7 million $\psi'$ events. BESII has collected 58
million $J/\psi$ events and 14 million $\psi'$ events since 1998.

\begin{figure}[htbp]
\includegraphics[scale=0.2]{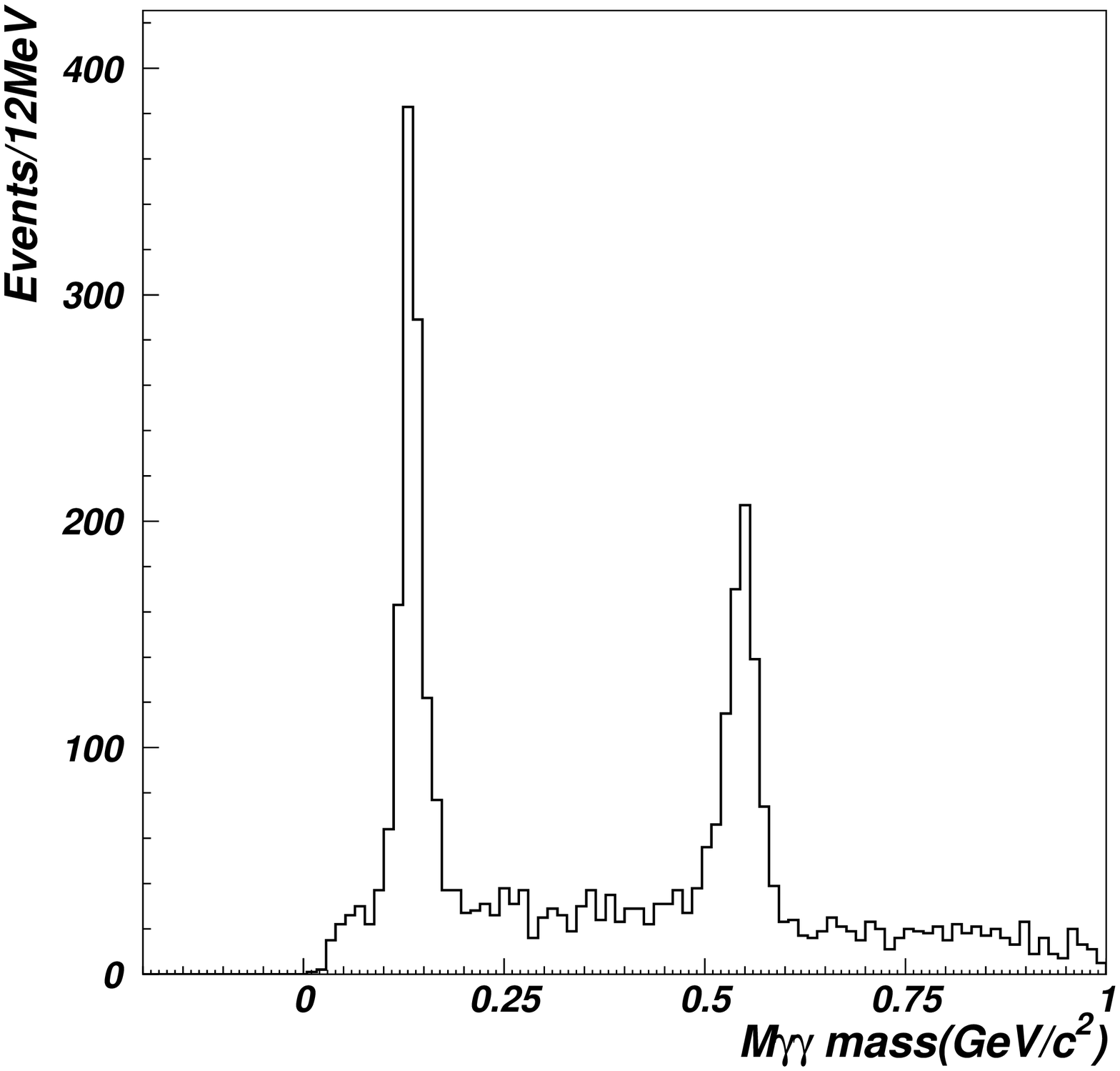}
\hspace{-0.5cm}\includegraphics[scale=0.2]{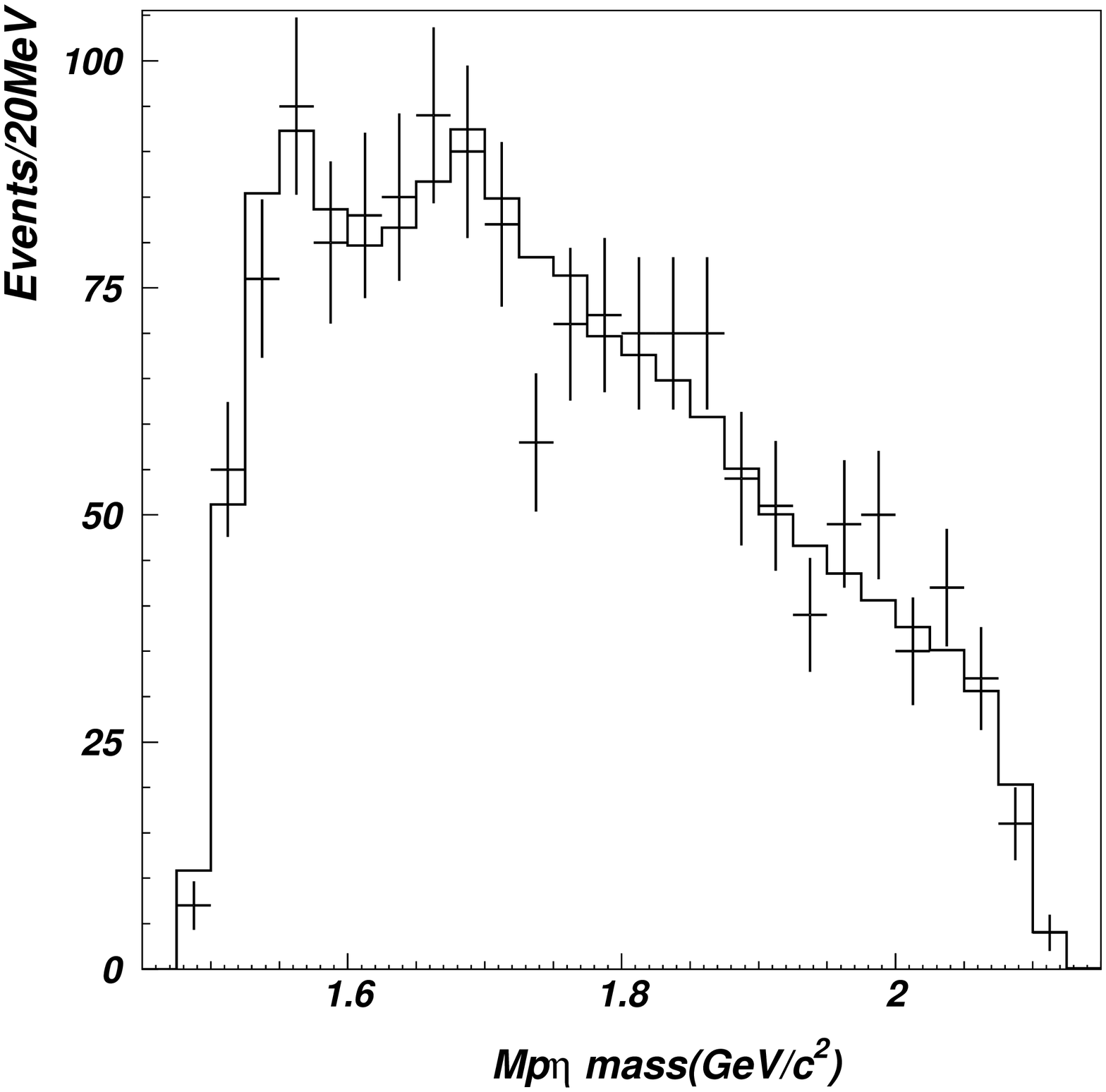}
\hspace{-0.5cm}\includegraphics[scale=0.2]{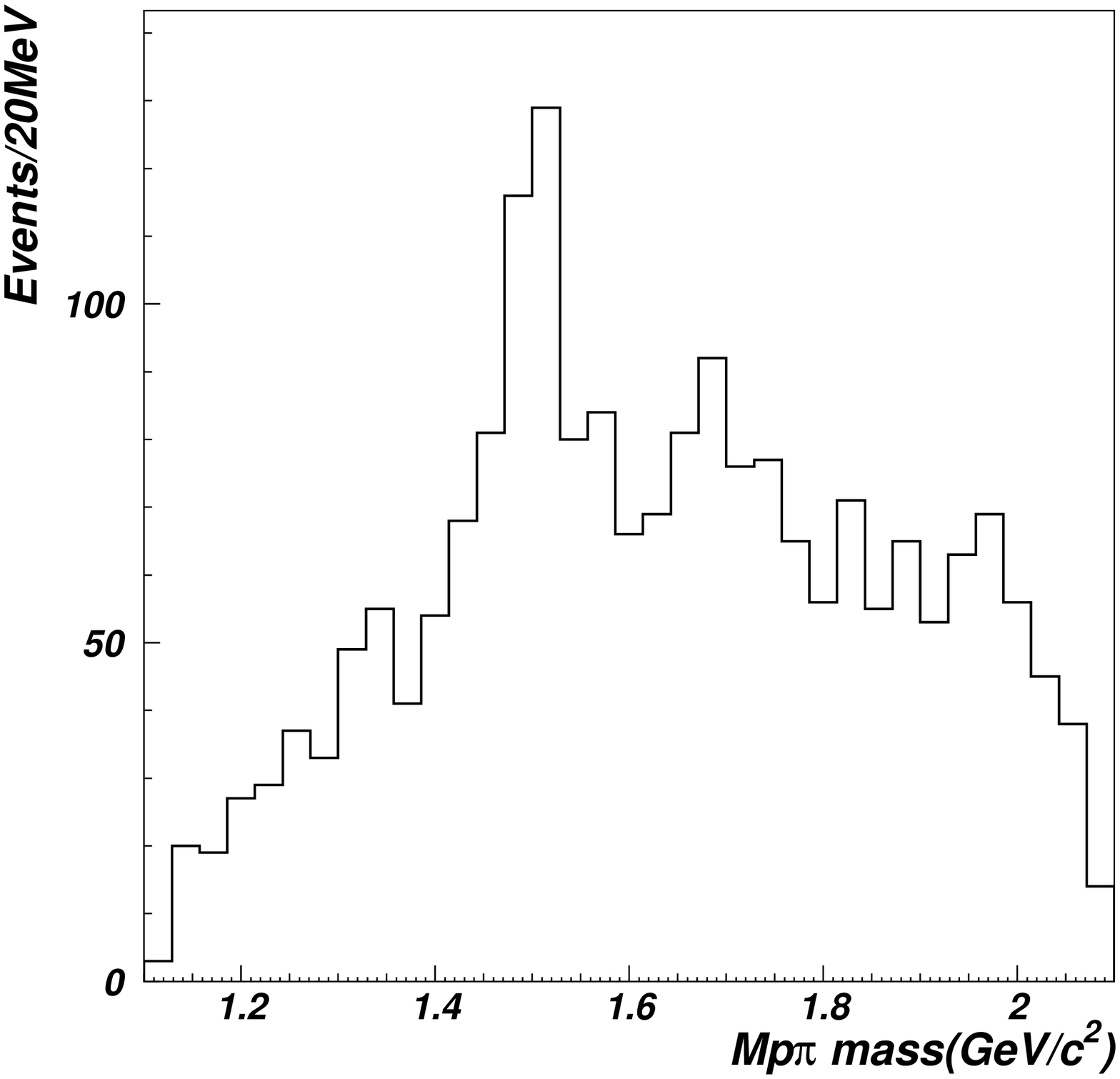}
\vspace{-0.5cm} \caption{\label{fig2} BESI data for $\gamma\gamma$
invariant mass of $J/\psi\to\bar pp\gamma\gamma$ (left); $p\eta$
(middle) and $p\pi$ (right) invariant mass spectra of $J/\psi\to\bar
pp\eta$ and $\bar pp\pi^0$. }
\end{figure}

Based on 7.8 million $J/\psi$ events collected at BESI from 1990 to
1991, the events for $J/\psi\to\bar pp\pi^0$ and $\bar pp\eta$ have
been selected and reconstructed with $\pi^0$ and $\eta$ detected in
their $\gamma\gamma$ decay mode \cite{Zou1}. The invariant mass of
$\gamma\gamma$ is shown in Fig.~\ref{fig2} (left) with two clear
peaks corresponding to $\pi^0$ and $\eta$. The $p\eta$ invariant
mass spectrum is shown in Fig.~\ref{fig2} (middle) with two peaks at
1540 and 1650 MeV. Partial wave analysis has been performed for the
$J/\psi\to\bar pp\eta$ channel \cite{Zou1} using the effective
Lagrangian approach \cite{Nimai,Olsson} with Rarita-Schwinger
formalism \cite{Rarita,Fronsdal,Chung,Liang} and the extended
automatic Feynman Diagram Calculation (FDC) package \cite{Wang}.
There is a definite requirement for a $J^{P}=\frac{1}{2}^-$
component at $M = 1530\pm 10$ MeV with $\Gamma =95\pm 25$ MeV near
the $\eta N$ threshold. In addition, there is an obvious resonance
around 1650 MeV with $J^P=\frac{1}{2}^-$ preferred, $M = 1647\pm 20$
MeV and $\Gamma = 145^{+80}_{-45}$ MeV. These two $N^*$ resonances
are believed to be the two well established states, $S_{11}(1535)$
and $S_{11}(1650)$, respectively. In the higher
$p\eta$($\bar{p}\eta$) mass region, there is an evidence for a
structure around 1800 MeV; with BESI statistics one cannot determine
its quantum numbers. The $p\pi^0$ invariant mass spectrum from
$J/\psi\to p\bar p\pi^0$ is shown in Fig.~\ref{fig2} (right) with
two clear peaks around 1500 and 1670 MeV, and some weak structure
around 2 GeV.

\begin{figure}[htbp]
\vspace{-0.7cm}
\includegraphics[scale=0.33]{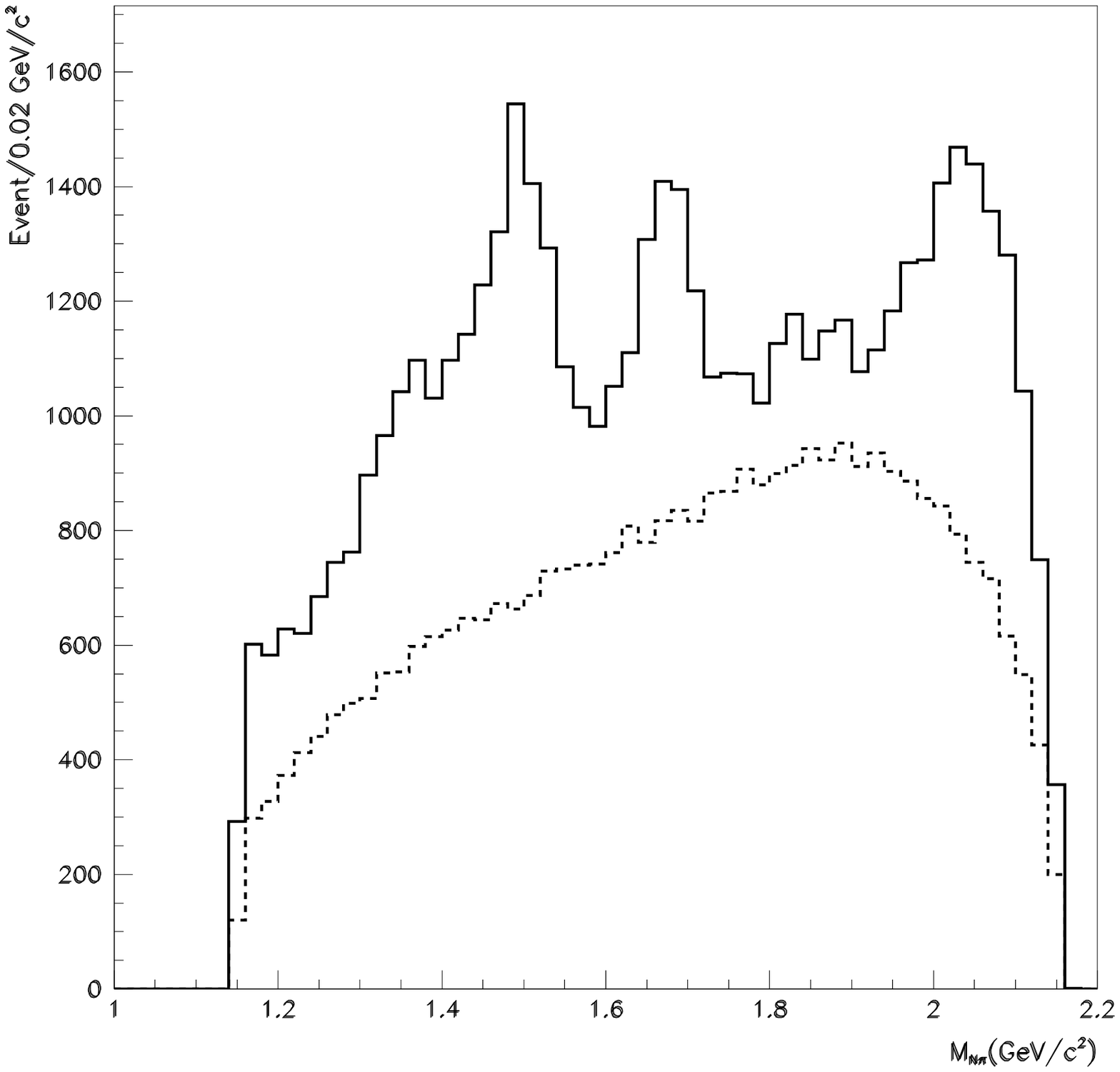}
\includegraphics[scale=0.6]{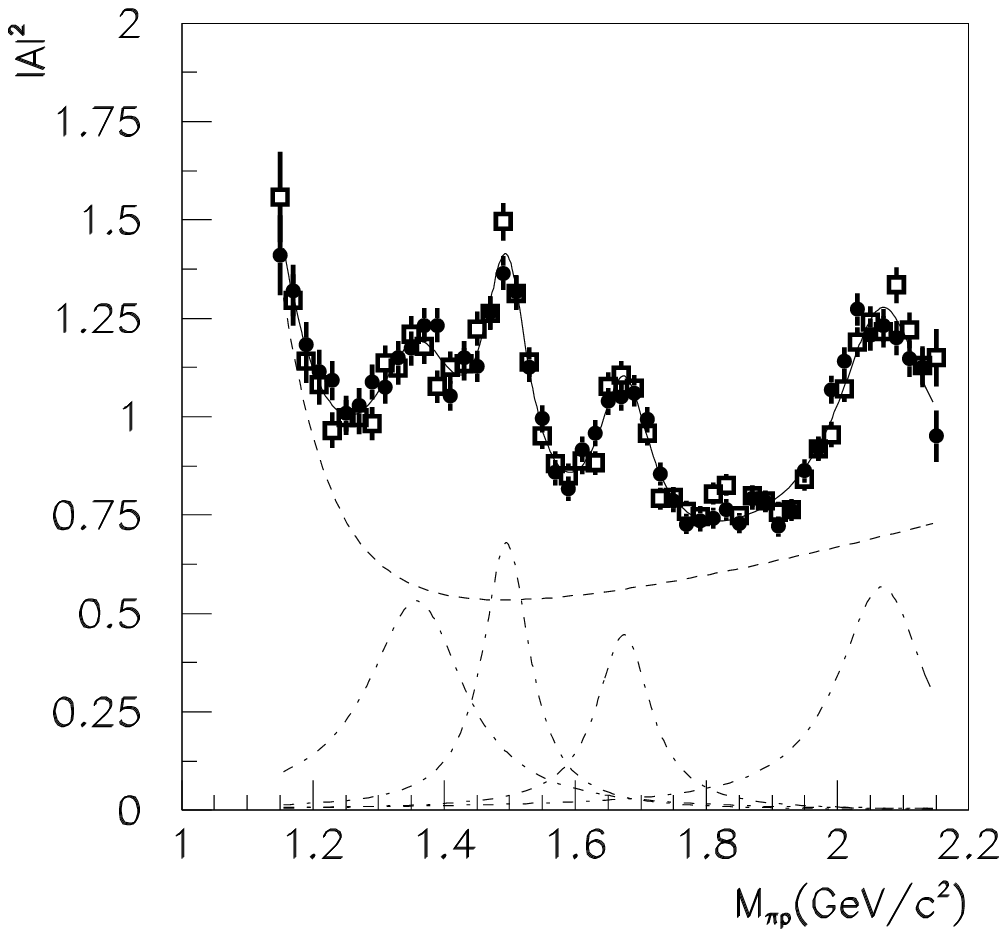}\vspace{-0.7cm}
\caption{\label{fig3.2} $p\pi^-$ invariant mass spectrum for
$J/\psi\to p\pi^-\bar n$ compared with phase space distribution
(left); And data divided by MC phase space vs $p\pi$ invariant mass
for $J/\psi\to \bar p\pi^-\bar n$ (solid circle) and $\bar p\pi^+n$
(open square).}
\end{figure}

\begin{figure}[htbp]
\includegraphics[scale=0.225]{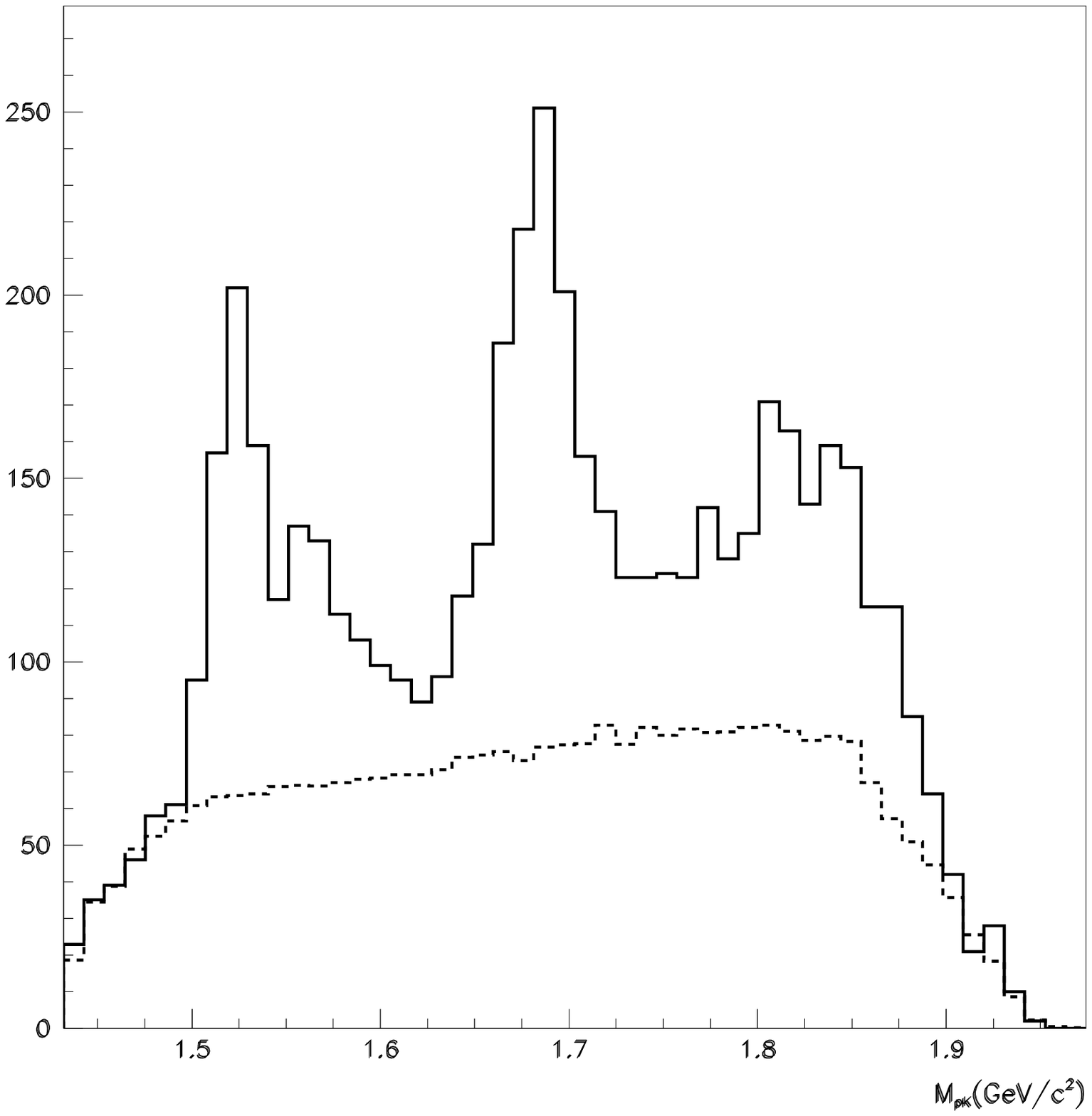}
\includegraphics[scale=0.225]{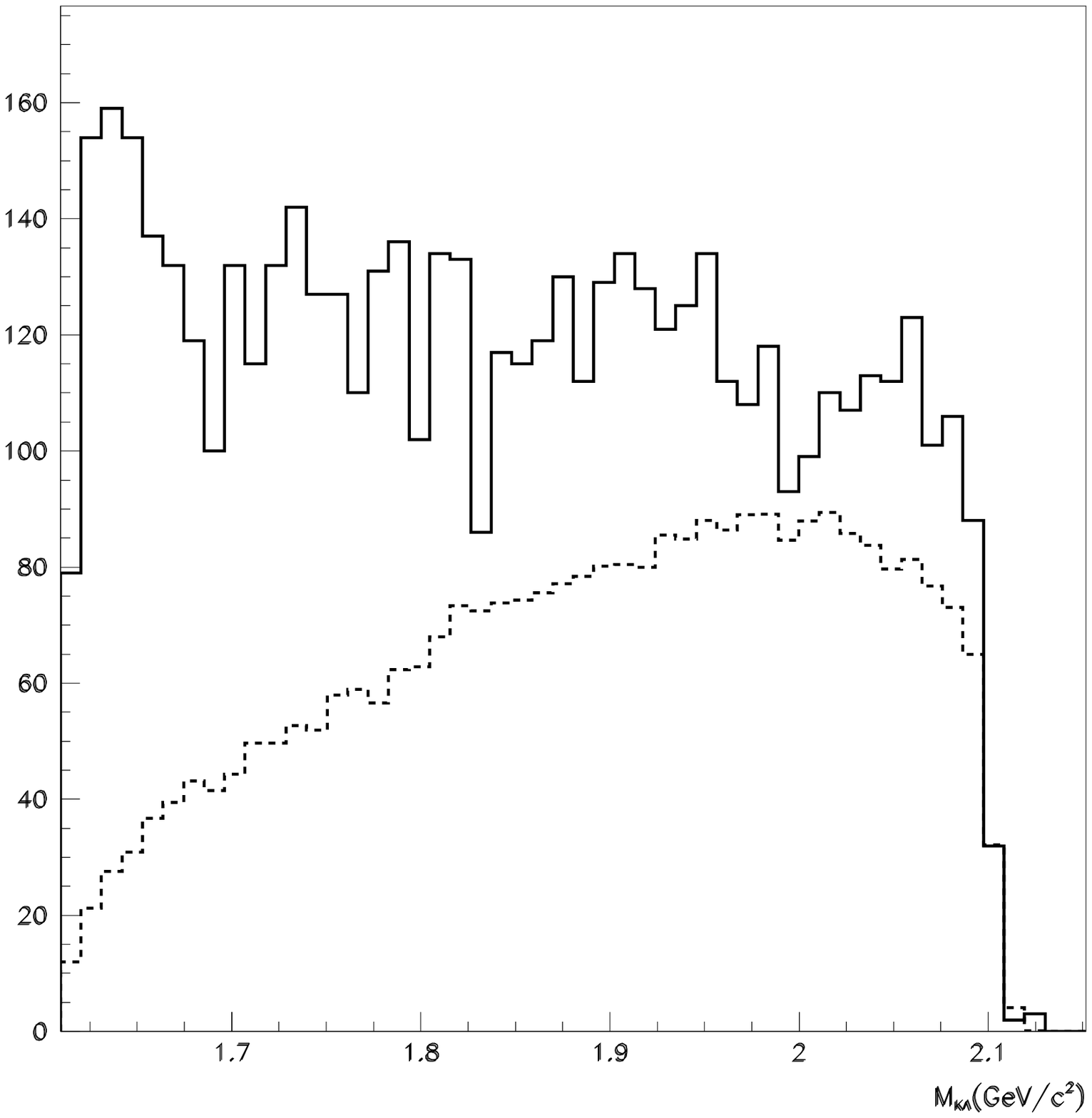}
\includegraphics[scale=0.225]{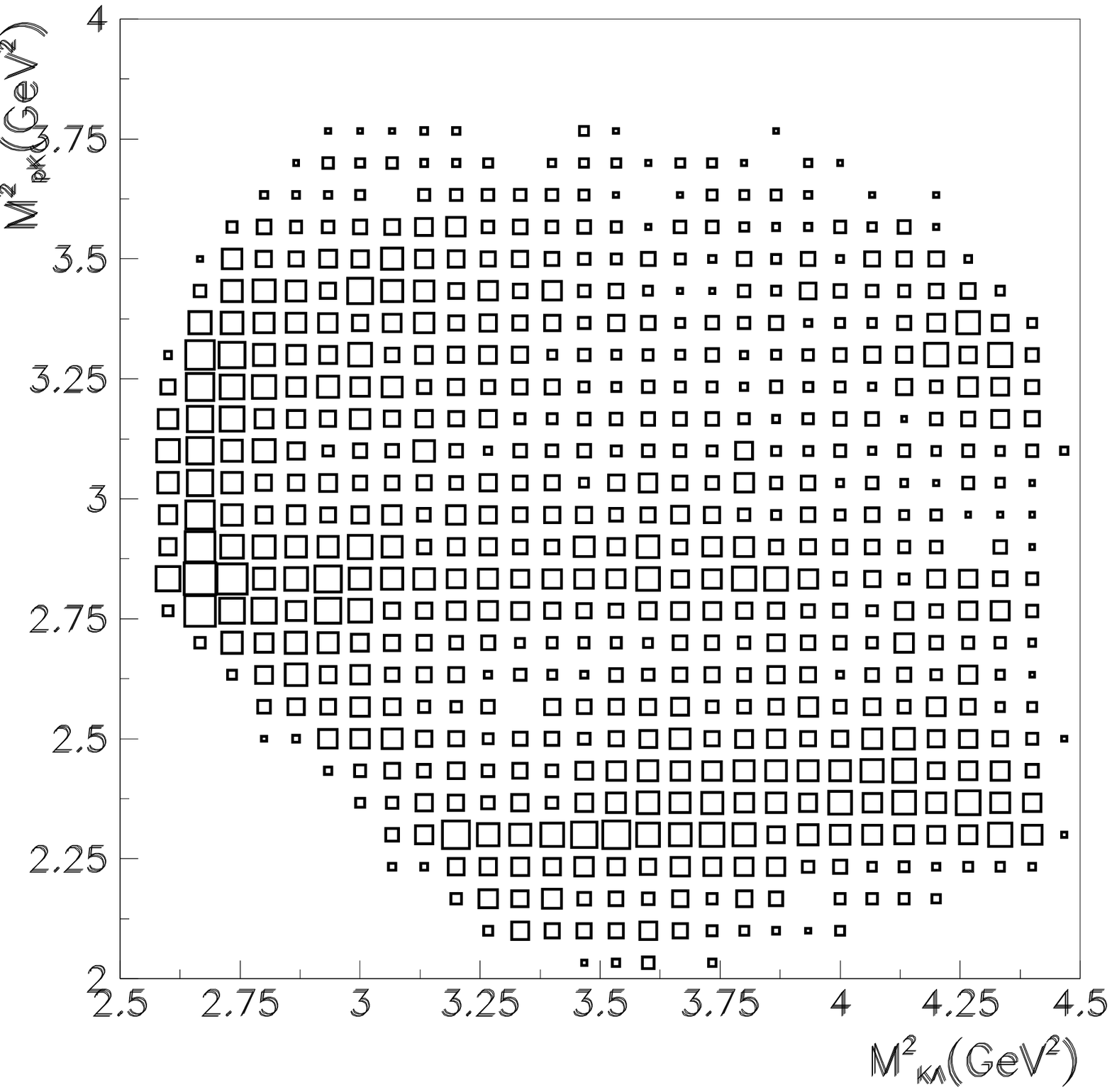}
\caption{\label{fig4} $pK$ (left) and $K\Lambda$ (middle) invariant
mass spectra for $J/\psi\to pK^-\bar\Lambda$+c.c., compared with
phase space distribution; right: Dalitz plot for $J/\psi\to
pK^-\bar\Lambda$+c.c.}
\end{figure}

\begin{figure}[htbp]
\begin{center}
\includegraphics[scale=0.27]{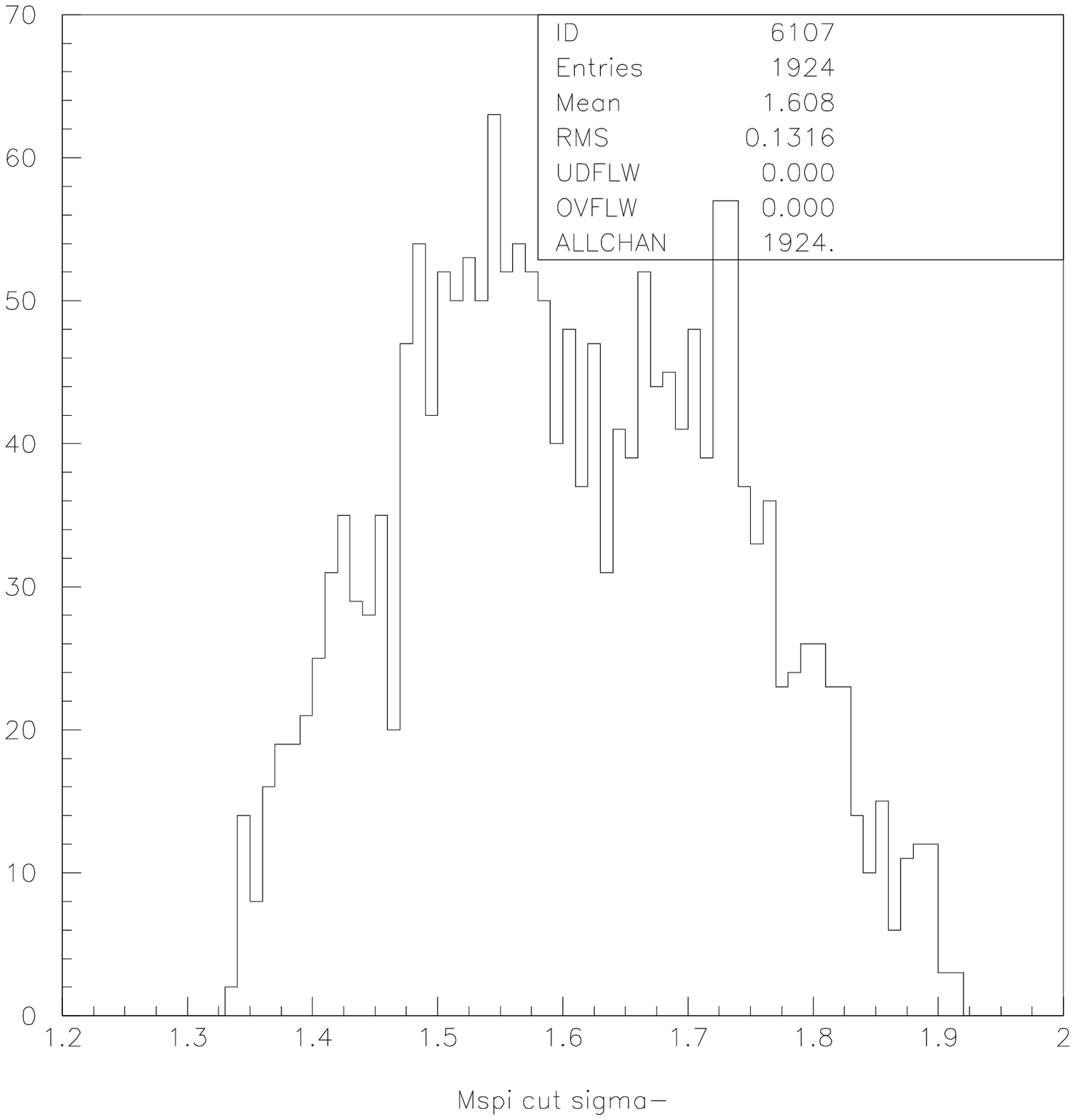}
\includegraphics[scale=0.27]{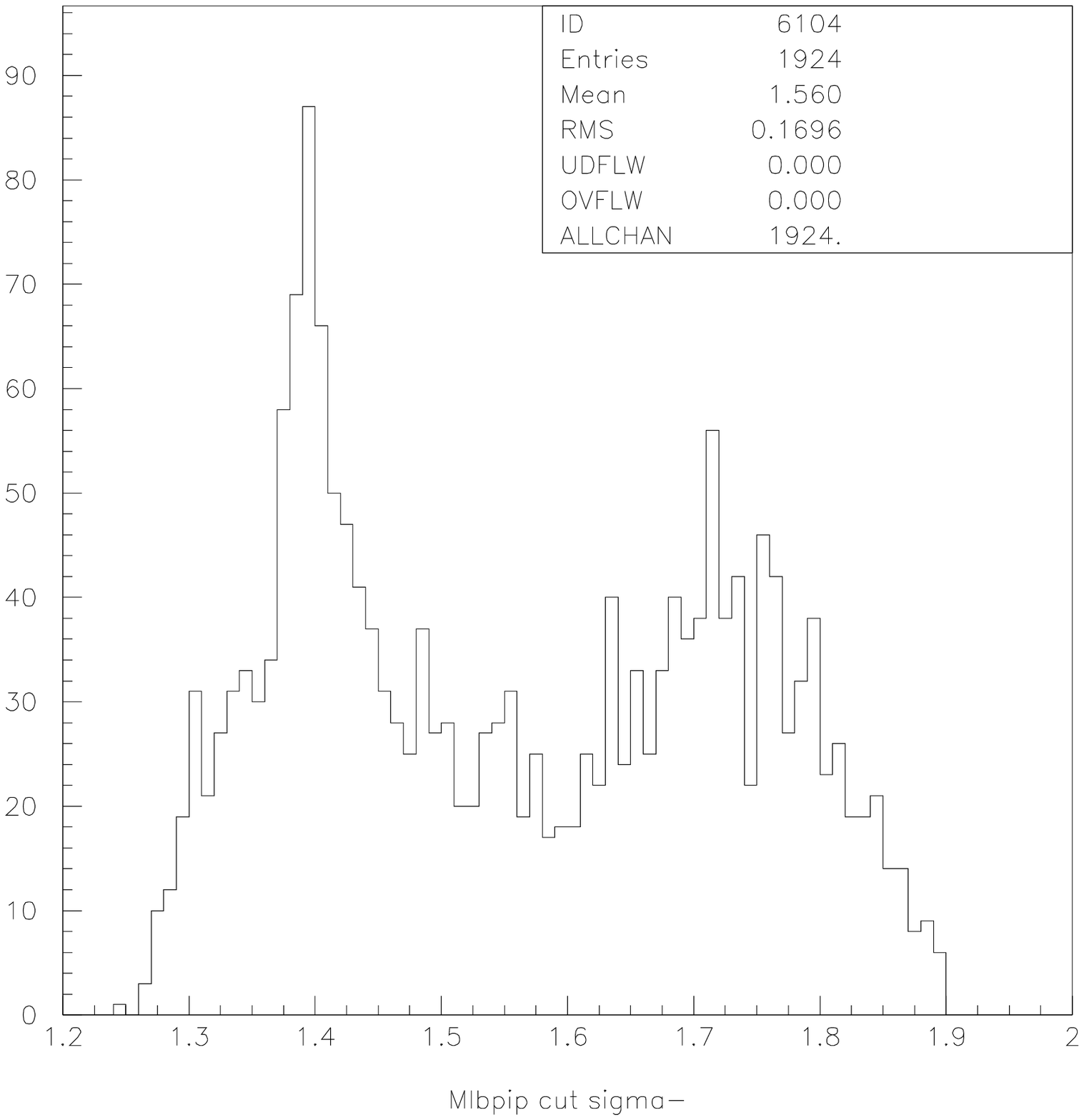}
\end{center}
\caption{\label{fig5} $\bar\Sigma\pi$ (left) and $\Lambda\pi$
(right) invariant mass spectrum for $J/\psi\to
\Lambda\bar\Sigma^+\pi^-$. Preliminary BESII data \cite{zoubs4}}
\end{figure}

With 58 million new $J/\psi$ events collected by BESII of improved
detecting efficiency, one order of magnitude more reconstructed
events can be obtained for each channel. Results for $J/\psi$ to
$p\bar n\pi^-+c.c.$, $pK^-\bar\Lambda + c.c.$ and
$\Lambda\bar\Sigma\pi$+c.c. channels are shown in
Figs.\ref{fig3.2},\ref{fig4},\ref{fig5}, respectively. These are
typical channels for studying $N^*$, $\Lambda^*$ and $\Sigma^*$
resonances. For $J/\psi\to p\bar p\pi^0$ channel, the $N\pi$
invariant mass spectrum looks similar to the BESI data, but with
much higher statistics.

For $J/\psi\to p\bar n\pi^-$ channel, proton and $\pi^-$ are
detected \cite{bes2}. With some cuts of backgrounds, the missing
mass spectrum shows a very clean peak for the missing antineutron.
In the $p\pi^-$ invariant mass spectrum as shown in Fig.\ref{fig3.2}
(left), besides two well known $N^*$ peaks at 1500 and 1670 MeV,
there are two new clear $N^*$ peaks around 1360 and 2030 MeV. Its
charge conjugate channel $\bar p\pi^+n$ gives very similar results.

To investigate the behavior of the amplitude squared as a function
of invariant mass, one should remove the phase space factor and
efficiency factor from the invariant mass distribution by dividing
the data by Monte Carlo phase space times the detection efficiency.
The results are shown in Fig.~\ref{fig3.2} (right). At low $p\pi$
invariant mass, the tail from nucleon pole term, expected from
theoretical considerations \cite{Okubo,Liang2}, is clearly seen.
There are clearly four peaks around 1360 MeV, 1500 MeV, 1670 MeV and
2065 MeV. Note that the well known first resonance peak
($\Delta(1232)$) in $\pi N$ and $\gamma N$ scattering data does not
show up here due to the isospin filter effect of our $J/\psi$ decay.
While the two peaks around 1500 MeV and 1670 MeV correspond to the
well known second and third resonance peaks observed in $\pi N$ and
$\gamma N$ scattering data, the two peaks around 1360 MeV and 2065
MeV have never been observed in $\pi N$ invariant mass spectra
before. The one around 1360 MeV should be from $N^*(1440)$ MeV which
has a pole around 1360 MeV \cite{PDG,Manley,Dytman} and which is
usually buried by the strong $\Delta$ peak in $\pi N$ and $\gamma N$
experiments; the other one around 2065 MeV may be due to the long
sought ``missing" $N^*$ resonance(s). For the decay $J/\psi\to\bar
NN^*(2065)$, the orbital angular momentum of $L=0$ is much preferred
due to the suppression of the centrifugal barrier factor for $L\geq
1$. For $L=0$, the spin-parity of $N^*(2065)$ is limited to be
$1/2+$ and $3/2+$. This may be the reason that the $N^*(2065)$ shows
up as a peak in $J/\psi$ decays while only much broader structures
show up for $\pi N$ invariant mass spectra above 2 GeV in $\pi N$
and $\gamma N$ production processes \cite{Gaohy} which allow all
$1/2\pm$, $3/2\pm$, $5/2\pm$ and $7/2\pm$ $N^*$ resonances around
2.05 GeV to overlap and interfere with each other there. A simple
Breit-Wigner fit \cite{bes2} gives the mass and width for the
$N^*(1440)$ peak as $1358\pm 6 \pm 16$ MeV and $179\pm 26\pm 50$
MeV. Very recently, CELSIUS-WASA Collaboration \cite{Clement} also
observed the $N^*(1440)$ peak in the $n\pi^+$ invariant mass
spectrum for their $pp\to pn\pi^+$ reaction and obtained mass and
width consistent with ours. For the new $N^*$ peak above 2 GeV the
fitted mass and width are $2068\pm 3^{+15}_{-40}$ MeV and $165\pm
14\pm 40$ MeV, respectively. A partial wave analysis indicates that
the $N^*(2065)$ peak contains both spin-parity $1/2+$ and $3/2+$
components \cite{bes2}.

For $J/\psi\to pK^-\bar\Lambda$ and $\bar pK^+\Lambda$ channels
\cite{yang}, there are clear $\Lambda^*$ peaks at 1.52 GeV, 1.69 GeV
and 1.8 GeV in $pK$ invariant mass spectrum, and $N^*$ peaks near
$K\Lambda$ threshold, 1.9 GeV  and 2.05 GeV for $K\Lambda$ invariant
mass spectrum. The $N^*$ peak near $K\Lambda$ threshold is most
probably due to $N^*(1535)$ which was found to have large coupling
to $K\Lambda$ \cite{Oset1,liubc}. The SAPHIR experiment at
ELSA\cite{ELSA} also observed a $N^*$ peak around 1.9 GeV for
$K\Lambda$ invariant mass spectrum from photo-production, and the
fit \cite{Mosel} to the data reveals large $1/2^-$ near-threshold
enhancement mainly due to the $N^*(1535)$. The $N^*$ peak at 2.05
GeV is compatible with that observed in $N\bar N\pi$ channels.

For $J/\psi\to\Lambda\Sigma\pi$ channels \cite{zoubs4}, it seems
also $\Lambda^*$ peaks at 1.52 GeV, 1.69 GeV and 1.8 GeV in
$\Sigma\pi$ invariant mass spectra, similar to those in the
$pK\Lambda$ channel, although less clear. In $\Lambda\pi$ invariant
mass spectra, there is a very clear peak around 1.385 GeV
corresponding to the well-established $\Sigma(1385)$ resonance and
there is also another $\Sigma^*$ peak around 1.72 GeV.

In order to get more useful information about properties of the
baryon resonances involved, such as their $J^{PC}$ quantum numbers,
mass, width, production and decay rates, etc., partial wave analysis
(PWA) is necessary. We use event-based standard maximum likelihood
method with partial wave amplitudes constructed by the effective
Lagrangian approach\cite{Nimai,Olsson} with Rarita-Schwinger
formalism\cite{Rarita,Fronsdal,Chung,Liang}.

\section{Baryon spectroscopy Prospects  at BESIII}

Recently, empirical indications for a positive strangeness magnetic
moment and positive strangeness radius of the proton suggest that
the 5-quark components in baryons may be largely in colored diquark
cluster configurations rather than mainly in ``meson cloud''
configurations \cite{zou1,zr}. The diquark cluster picture also
gives a natural explanation for the excess of $\bar d$ over $\bar u$
in the proton with a mixture of $[ud][ud]\bar d$ component in the
proton. More precise measurements and analyses of the strange form
factors are needed to examine the relative importance of the
meson-cloud components and $q^2q^2\bar q$ components in the proton.

For excited baryons, the excitation energy for a spatial excitation
could be larger than to drag out a $q\bar q$ pair from gluon field
with the $q$ to form diquark cluster with a valence quark. Hence the
5-quark components could be dominant for some excited baryons.

The diquark cluster picture for the 5-quark components in baryons
also gives a natural explanation for the longstanding mass-reverse
problem of $N^*(1535)$, $N^*(1440)$ and $\Lambda^*(1405)$ resonances
as well as the unusual decay pattern of the $N^*(1535)$ resonance
with a large $|[ud][us]\bar s>$ component \cite{zou1,liubc}.

The diquark cluster picture predicts the existence of the SU(3)
partners of the $N^*(1535)$ and $\Lambda^*(1405)$, {\sl i.e.}, an
additional $\Lambda^*~1/2^-$ around 1570 MeV, a triplet
$\Sigma^*~1/2^-$ around 1360 MeV and a doublet $\Xi^*~1/2^-$ around
1520 MeV \cite{zhusl}. Although there is no observation of these
resonances~\cite{PDG}, they may hide underneath the peaks of
$\Lambda^*(1600)$, $\Sigma^*(1385)$ and $\Xi^*(1530)$, respectively.

According to PDG \cite{PDG}, the branching ratios for
$J/\psi\to\bar\Sigma^-\Sigma^*(1385)^+$ and
$J/\psi\to\bar\Xi^+\Xi^*(1530)^-$ are $(3.1\pm 0.5)\times 10^{-4}$
and $(5.9\pm 1.5)\times 10^{-4}$, respectively. These two processes
are SU(3) breaking decays since $\Sigma$ and $\Xi$ belong to SU(3)
$1/2^+$ octet while $\Sigma^*(1385)$ and $\Xi^*(1530)$ belong to
SU(3) $3/2^+$ decuplet. Comparing with the similar SU(3) breaking
decay $J/\psi\to\bar p\Delta^+$ with branching ratio of less than
$1\times 10^{-4}$ and the SU(3) conserved decay $J/\psi\to\bar
pN^*(1535)^+$ with branching ratio of $(10\pm 3)\times 10^{-4}$, the
branching ratios for $J/\psi\to\bar\Sigma^-\Sigma^*(1385)^+$ and
$J/\psi\to\bar\Xi^+\Xi^*(1530)^-$ are puzzling too high. A possible
explanation for this puzzling phenomena is that there were
substantial components of $1/2^-$ under the $3/2^+$ peaks but the
two branching ratios were obtained by assuming pure $3/2^+$
contribution. This possibility should be easily checked with the
high statistics BESIII data in near future.

\begin{figure}[htbp]
\includegraphics[scale=0.25]{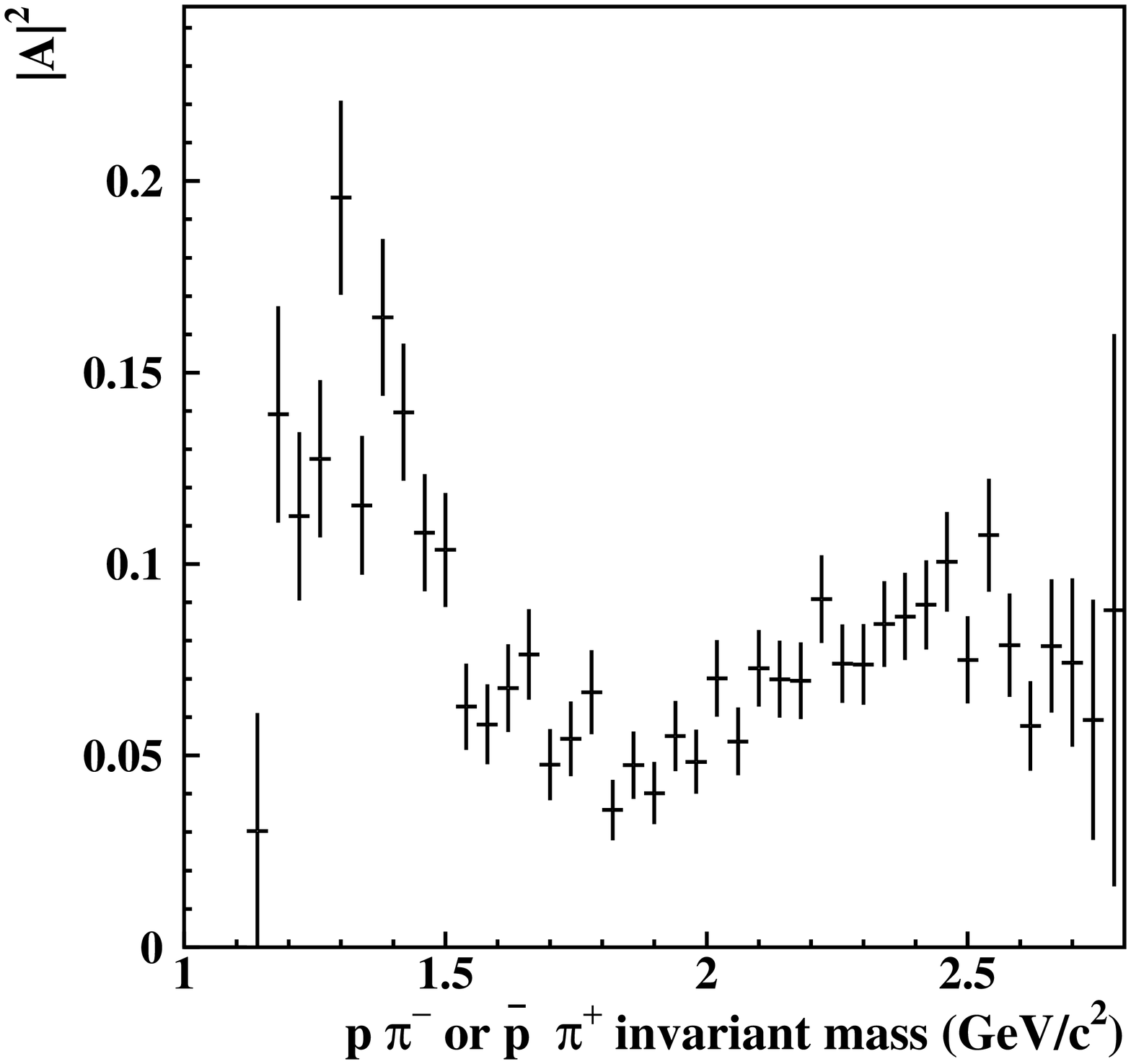}
\includegraphics[scale=0.25]{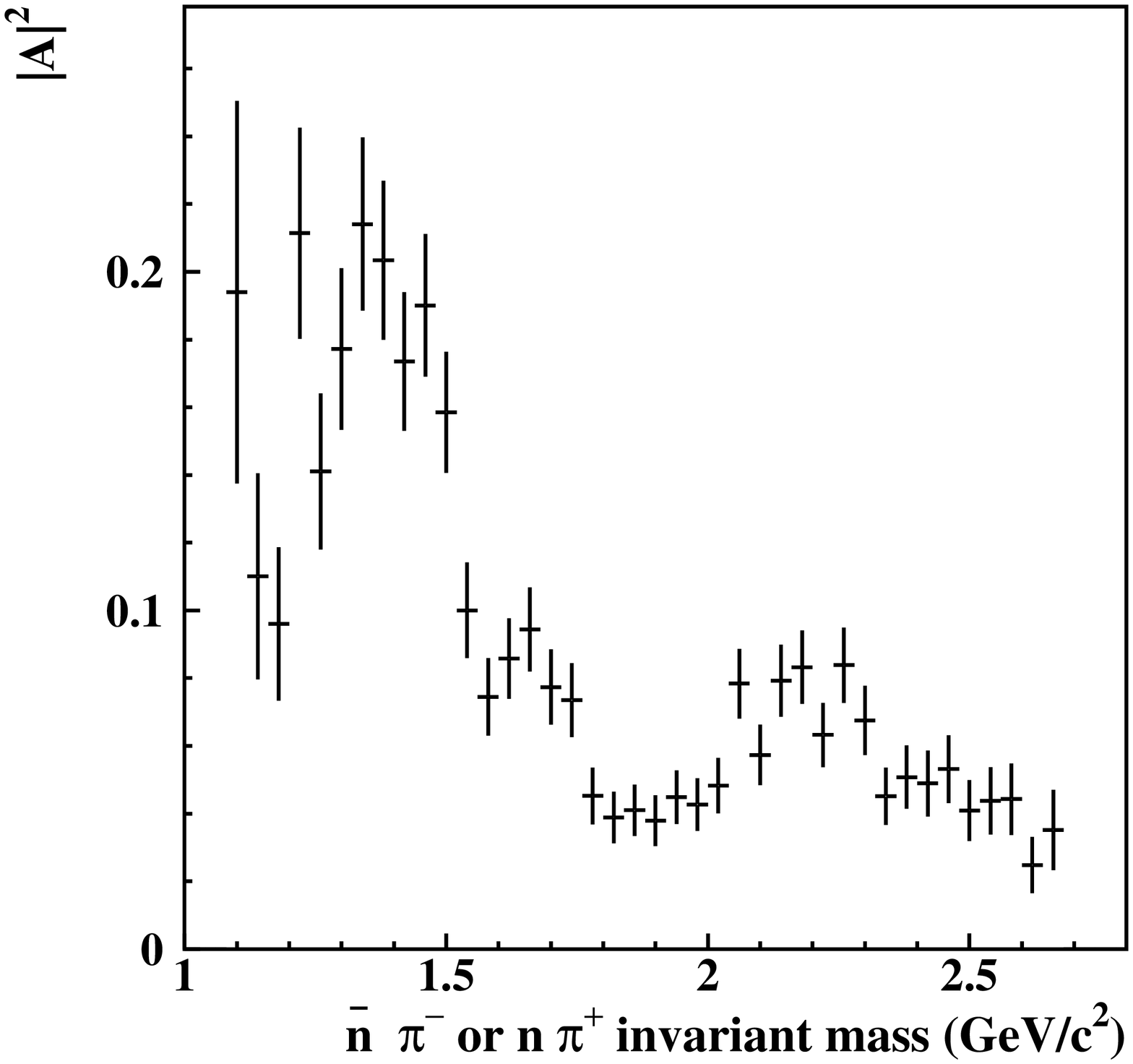}
\caption{\label{am} Data divided by efficiency and phase space vs $p
\pi^-$ (or $\bar{p} \pi^+$) and $\bar{n} \pi^-$ (or $n \pi^+$)
invariant mass for $\psi' \to p \bar{n} \pi^-+c.c.$ candidate events
\cite{psip2}.}
\end{figure}

With two order of magnitude more statistics at BESIII, plenty
important channels for baryon spectroscopy can be studied from both
$J/\psi$ and $\psi'$ decays. The $\psi'$ data will significantly
extend the mass range for the study of baryon spectroscopy. For
example, for $\psi'\to p \bar{n} \pi^-+c.c.$ events collected at
BESII \cite{psip2}, there are obvious structures for $M_{N\pi}>
2$~GeV in the $N\pi$ invariant mass spectra as shown in
Fig.~\ref{am}. However due to low statistics at BESII, no conclusive
information can be drawn for the $N^*$ resonances with mass above 2
GeV from $\psi'$ decays \cite{psip2,psip1}. With BESIII statistics,
determination of properties for these high mass $N^*$ resonances can
be done. The BESIII $\psi'$ data will enable us to complete the
$\Lambda^*$, $\Sigma^*$ and $\Xi^*$ spectrum and examine various
pictures for their internal structures, such as simple 3q quark
structure and more complicated structure with pentaquark components
dominated.

\begin{table}[htb]
\caption{ Measured $J/\psi$ decay branching ratios (BR$\times 10^3$)
for channels involving baryon anti-baryon and meson(s)
\cite{PDG,bes2}} \label{table:1}
\renewcommand{\arraystretch}{1.2} 
\begin{tabular}{cccccc}
\hline $p\bar n\pi^-$  & $p\bar p\pi^0$  &  $p\bar p\pi^+\pi^-$ &
$p\bar p\eta$ & $p\bar p\eta'$ & $p\bar p\omega$ \\
\hline $2.4\pm 0.2$ & $1.1\pm 0.1$ & $6.0\pm 0.5$ & $2.1\pm 0.2$ &
$0.9\pm 0.4$ & $1.3\pm 0.3$ \\
\hline $\Lambda\bar\Sigma^-\pi^+$ & $pK^-\bar\Lambda$ &
$pK^-\bar\Sigma^0$
& $\bar pp\phi$ & $\Delta(1232)^{++}\bar p\pi^-$ & $pK^-\bar\Sigma(1385)^0$ \\
\hline $1.1\pm 0.1$ & $0.9\pm 0.2$ & $0.3\pm 0.1$
& $0.045\pm 0.015$ & $1.6\pm 0.5$ & $0.51\pm 0.32$ \\
\hline
\end{tabular}\\
\end{table}

The measured $J/\psi$ decay branching ratios for channels involving
baryon anti-baryon plus meson(s) are listed in Table \ref{table:1}.
With $10^{10}$ $J/\psi$ events, all these channels will get enough
statistics for partial wave analysis. Among these channels, the
$\Sigma\bar\Lambda\pi +c.c.$ channels should have high priority for
pinning down the lowest $1/2^-$ $\Sigma^*$ and $\Lambda^*$ as well
as other higher excited $\Sigma^*$ and $\Lambda^*$ states. Another
very important channel is $K^-\Lambda\bar\Xi^+ + c.c.$ which is the
best channel for finding the lowest $1/2^-$ $\Xi^*$ resonance and
many other ``missing" $\Xi^*$ states with  $\Xi^*\to K\Lambda$. This
channel should be rather easy to be reconstructed by BESIII. One can
select events containing $K^-$ and $\Lambda$ with $\Lambda\to
p\pi^-$, then from missing mass spectrum of $K^-\Lambda$ one should
easily identify the very narrow $\bar\Xi^+$ peak.

For $10^9$ $\psi^\prime$ events, the $K^-\Lambda\bar\Xi^+ + c.c.$
and $p\bar p\phi$ channels should have high priority. These two
channels are strongly limited by phase space in $J/\psi$ decays.
From $\psi^\prime$ decays, the phase space is much increased. The
$K^-\Lambda\bar\Xi^+ + c.c.$ channel should allow us to discover
many ``missing" $\Xi^*$ resonances, while the $p\bar p\phi$ channel
should allow us to find those $N^*$ resonances with large coupling
to $N\phi$ \cite{huangf} and hence large 5-quark components.

After analyzing the easier 3-body final states, 4-body and 5-body
channels should also be investigated. Among them,
$\Delta(1232)^{++}\bar p\pi^-$ in $p\bar p\pi^+\pi^-$ and
$\Delta(1232)^{++}\bar\Sigma^-K^-$ in $p\bar\Sigma^-\pi^+K^-$ are
very good channels for finding ``missing" $\bar\Delta^{*--}$
decaying to $\bar p\pi^-$ and $\bar\Sigma^-K^-$, respectively. The
spectrum of isospin 3/2 $\Delta^{++*}$ resonances is of special
interest since it is the most experimentally accessible system
composed of 3 identical valence quarks. Recently, the lowest $1/2^-$
baryon decuplet is proposed to contain large vector-meson-baryon
molecular components \cite{xiejj}. In the new scheme, the
$\Xi^*(1950)$ is predicted to be $1/2-$ resonance with large
coupling to $\Lambda K^*$. The $\psi^\prime\to\bar\Xi\Lambda K^*$
will provide a very good place to look for ``missing" $\Xi^*$ with
large coupling to $\Lambda K^*$.

In summary, BESIII data can play a very important role in studying
excited nucleons and hyperons, {i.e.}, $N^*$, $\Lambda^*$,
$\Sigma^*$, $\Xi^*$ and $\Delta^{*++}$ resonances.

\bigskip
\noindent {\bf Acknowledgements : } I would like to thank my BES
colleagues for producing nice results presented here. This work is
partly supported by the National Natural Science Foundation of China
under grants Nos. 10435080, 10521003 and by the CAS under project
No. KJCX3-SYW-N2.

\end{document}

%% file: MENU07macros.tex
\newcommand{\beq}{\begin{equation}}
\newcommand{\eeq}[1]{\label{#1}\end{equation}}
\newcommand{\eeqn}{\end{equation}}

\newcommand{\beqa}{\begin{eqnarray}}
\newcommand{\eeqa}[1]{\label{#1}\end{eqnarray}}
\newcommand{\eeqan}{\end{eqnarray}}

\let\bar=\overbar

\newcommand{\Dslash}{\not{\hbox{\kern-4pt $D$}}}
\newcommand{\dslash}{\not{\hbox{\kern-2pt $\del$}}}

\newcommand{\msb}{{\bar{\ssstyle M \kern -1pt S}}}

